\documentclass[aps,prl,twocolumn,superscriptaddress,showpacs]{revtex4-1}
\usepackage[utf8]{inputenc}
\usepackage{amsmath}
\usepackage{amssymb}
\usepackage{latexsym}
\usepackage{ifpdf}

\ifpdf
\usepackage[pdftex]{graphicx}
\else
\usepackage{graphicx}
\fi

\ifpdf
\DeclareGraphicsExtensions{.pdf, .jpg, .tif}
\else
\DeclareGraphicsExtensions{.eps, .jpg}
\fi

\begin{document}

\title{Enhanced radiative heat transfer between nanostructured gold plates}

\author{R. Guérout}
\affiliation{Laboratoire Kastler-Brossel, CNRS, ENS, UPMC, Case 74, F-75252 Paris, France}
\author{J. Lussange}
\affiliation{Laboratoire Kastler-Brossel, CNRS, ENS, UPMC, Case 74, F-75252 Paris, France}
\author{F. S. S. Rosa}
\affiliation{Laboratoire Charles Fabry, Institut d'Optique, CNRS,
Université Paris-Sud, Campus Polytechnique, RD128, F-91127
Palaiseau Cedex, France}
\author{J.-P. Hugonin}
\affiliation{Laboratoire Charles Fabry, Institut d'Optique, CNRS,
Université Paris-Sud, Campus Polytechnique, RD128, F-91127
Palaiseau Cedex, France}
\author{D. A. R. Dalvit}
\affiliation{Theoretical Division, MS B213, Los Alamos National
Laboratory, Los Alamos, New Mexico 87545, USA}
\author{J.-J. Greffet}
\affiliation{Laboratoire Charles Fabry, Institut d'Optique, CNRS,
Université Paris-Sud, Campus Polytechnique, RD128, F-91127
Palaiseau Cedex, France}
\author{A. Lambrecht}
\affiliation{Laboratoire Kastler-Brossel, CNRS, ENS, UPMC, Case 74, F-75252 Paris, France}
\author{S. Reynaud}
\affiliation{Laboratoire Kastler-Brossel, CNRS, ENS, UPMC, Case 74, F-75252 Paris, France}

\date{\today}

\begin{abstract}
We compute the radiative heat transfer between nanostructured gold
plates in the framework of the scattering theory. We predict an
enhancement of the heat transfer as we increase the depth of the
corrugations while keeping the distance of closest approach fixed.
We interpret this effect in terms of the evolution of plasmonic and
guided modes as a function of the grating's geometry.
\end{abstract}


\maketitle

\label{ssub:introduction}

The far-field radiative heat transfer between good conductive metals
is very low at room temperature, since they are very good reflectors
at the infrared frequencies of blackbody radiation. The radiative
heat transfer is enhanced in the near field, due to the contribution
of evanescent surface modes~\cite{polder1971,volokitin2007,joulain2005}. 
Polar materials like SiO$_{2}$ or SiC
are in addition favored by the contribution of surface phonon
polaritons whose resonance frequencies lie in the
infrared~\cite{MuletOC}. There is an analogous effect for metals
arising from the surface plasmons resonances but those lie in the
ultraviolet and do not contribute significantly to the heat
transfer~\cite{shenNanoLetters2009}.

It has been shown recently that the radiative heat transfer can be
controlled by nanostructuring the interfaces periodically. When the
period $d$ is much smaller than the wavelength $\lambda$ and the
separation distance $L$, the system can be treated using an
effective refractive index for the equivalent homogeneous medium. It
has been shown that the induced anisotropy introduces additional
modes~\cite{BiehsOE} and also allows modulating the
flux~\cite{BiehsAPL}. For periods on the order of the wavelength, a
full solution of Maxwell equations is needed. The heat transfer
between two periodic slabs has been studied within a two dimensional
approximation and for p-polarization using a finite difference time
domain (FDTD) technique~\cite{RodriguezPRL2011}. A flux enhancement
attributed to the excitation of the structure's modes was found. While
FDTD allows modeling complex shapes easily, dealing with bulk 3D
media and accounting for polarization effects has not been achieved
so far.

In this letter, we compute the radiative heat transfer between 1D
gold lamellar gratings in the framework of the scattering theory. We
do include all propagation directions (the so-called conical
diffraction) and all polarization states, which is of critical
importance in order to deal quantitatively with cross-polarization
effects~\cite{marquierOE}. The scattering theory is the most
successful technique for treating the Casimir effect between bodies
at thermodynamic equilibrium~\cite{LambrechtCP2011,RahiCP2011}. The
method determines the electromagnetic field in the space between the
two bodies in interaction in order to compute the Maxwell stress
tensor in terms of the reflection amplitudes on the two bodies. When
the two bodies are not at the same temperature, there is a net flux
of energy transferred from the warm body to the cold one. Recently,
this heat transfer problem between two bodies kept at different
temperatures has also been formulated in terms of the scattering
properties of the 
bodies~\cite{bimonte2009,messina,krueger2011,BenAbdallah2011}.

In the following, we use the scattering amplitudes which have
already been calculated for studying the Casimir interaction between
1D lamellar gratings~\cite{lambrechtPRL2008} and deduce the heat
flux when the two bodies are at different temperatures. We show that
the heat flux is largely enhanced when the corrugation depth is
increased while keeping the distance of closest approach fixed. We
attribute the heat flux increase to the excitation of guided modes
and surface plasmons whose frequencies change with the corrugation
depth.

\begin{figure}[hbtp]
\begin{center}
\includegraphics[width=5cm]{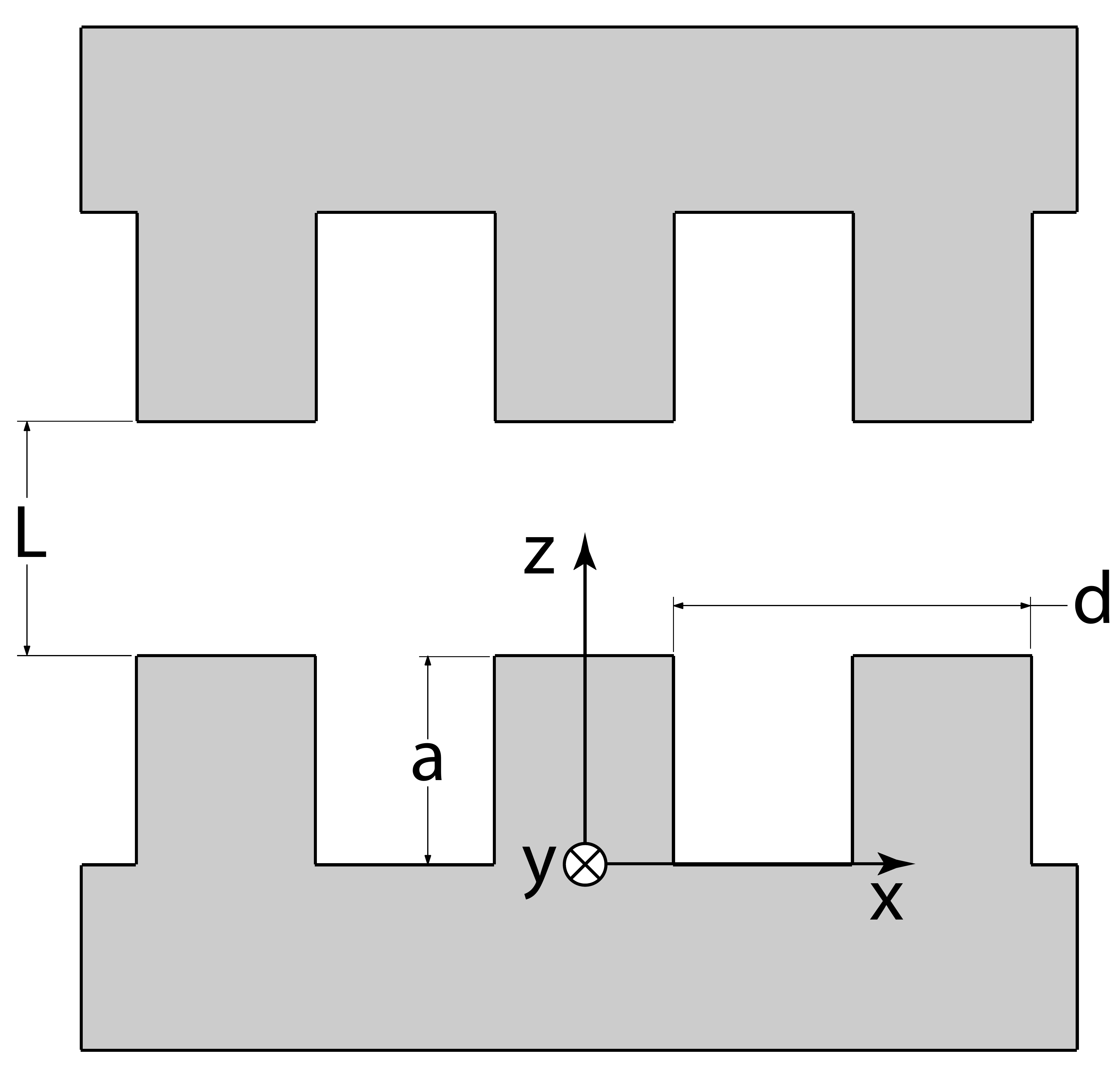}
\end{center}
\caption{The conventions used in the present paper. The grating
period is $d$, the corrugation depth is $a$, the distance of closest
approach of the two gratings is $L$. The lines of the grating are
along the $y$ direction, while the Fabry-Perot cavity between the
two gratings is along the $z$ direction.}
  \label{fig:gratingSchema}
\end{figure}

We consider the cavity formed by two gratings separated by a
distance of closest approach $L$ measured so as to vanish at contact
(Fig.~\ref{fig:gratingSchema}). The gratings are aligned and not
displaced laterally. We model the gold permittivity with a Drude
model
$\epsilon(\omega)=1-\frac{\omega_\mathrm{P}^{2}}{\omega(\omega+\imath
\gamma)}$ with $\omega_\mathrm{P}=9$ eV and $\gamma=35$ meV. We
write the heat flux $q$ between two bodies at temperatures $T_{1}$
and $T_{2}$ as \cite{joulain2005}
\begin{equation}
  \label{eq:heat}
q=\iiint\left(e_{T_1}(\omega)-e_{T_2}(\omega)\right)\mathcal{T}_L(\mathbf{k},\omega)
\,\frac{\mathrm{d}\omega\mathrm{d}^{2}\mathbf{k}}{(2\pi)^{3}},
\end{equation}
where
$e_T(\omega)=\hbar\omega\left(e^{\hbar\omega/k_{B}T}-1\right)^{-1}$
is the mean energy per mode of frequency $\omega$ at temperature $T$
while $\mathcal{T}_L(\mathbf{k},\omega)$ is the sum (trace) of the transmission 
factors for all the modes of frequency $\omega$ and lateral
wavevector $\mathbf{k}$ between the two gratings separated by a
distance $L$~\cite{biehs2010,pendry1983}. The expression of this 
transmission factor is given by
scattering amplitudes
\begin{subequations}
  \label{eq:Hs1s2}
    \begin{align}
      &\mathcal{T}_L(\mathbf{k},\omega)=
      \text{tr}\left(\mathbf{D}\mathbf{W_{1}}\mathbf{D}^{\dagger}\mathbf{W_{2}}\right),\\
      &\mathbf{D}=(\mathbf{1}-\mathbf{S_{1}}\mathbf{S_{2}})^{-1},\\
      \mathbf{W_{1}}&=\mathbf{\Sigma}_{-1}^{pw}-\mathbf{S_{1}}\mathbf{\Sigma}_{-1}^{pw}\mathbf{S_{1}}^{\dagger}
      +\mathbf{S_{1}}\mathbf{\Sigma}_{-1}^{ew}-\mathbf{\Sigma}_{-1}^{ew}\mathbf{S_{1}}^{\dagger},\\
      \mathbf{W_{2}}&=\mathbf{\Sigma}_{1}^{pw}-\mathbf{S_{2}}^{\dagger}\mathbf{\Sigma}_{1}^{pw}\mathbf{S_{2}}
      +\mathbf{S_{2}}^{\dagger}\mathbf{\Sigma}_{1}^{ew}-\mathbf{\Sigma}_{1}^{ew}\mathbf{S_{2}},\\
      &\mathbf{S_{1}}=\mathbf{R_{1}}(\mathbf{k},\omega),\\
      &\mathbf{S_{2}}=e^{\imath k_{z}L}\mathbf{R_{2}}(\mathbf{k},\omega)e^{\imath
      k_{z}L}.
    \end{align}
\end{subequations}
Mode counting is defined over frequency $\omega$ and lateral
wavevector $\mathbf{k}$ restricted to the first Brillouin zone, due
to the Bloch theorem.
$k_{z}=\sqrt{\omega^{2}/c^2-\mathbf{k}^{2}}$ is the
longitudinal wavevector for the Fabry-Perot cavity, with the
principal square root used in its definition
$-\frac{\pi}{2}<\text{arg}\,k_{z}\leq\frac{\pi}{2}$. The operators
$\mathbf{\Sigma}_{n}^{pw/ew}=k_{z}^{n}\mathbf{\Pi}^{pw/ew}$ involve
the projectors $\mathbf{\Pi}^{pw/ew}$ on the propagative or the
evanescent sector, respectively. $\mathbf{S_{1}}$ and
$\mathbf{S_{2}}$ are scattering operators defined from the
reflection operators $\mathbf{R_{1}}(\mathbf{k},\omega)$ and
$\mathbf{R_{2}}(\mathbf{k},\omega)$.
$\mathbf{S_{i}}$ are represented in the basis of the wavevectors
$\{\mathbf{k}^{(n)}\}$ coupled by the grating. We define
$\mathbf{k}^{(n)}=\mathbf{k}+n\frac{2\pi}{d}\hat{\mathbf{e}}_{x}$
where $d$ is the grating period, $\hat{\mathbf{e}}_{x}$ the
direction perpendicular to the lines of the grating (see
figure~\ref{fig:gratingSchema}) and $n$ runs from $-N$ to $+N$,
where $N$ is the highest diffraction order retained. The operators
$\mathbf{S_{i}}$ are square matrices of dimension
$2(2N+1)$~\cite{lambrechtPRL2008} as well as all bold operators
appearing in eqs.~\ref{eq:Hs1s2}. All scattering operators appearing
in eqs.~\ref{eq:Hs1s2} are represented in the $(s/p)$ (also denoted
TE/TM) polarization basis, well adapted to propagative fields. The reflection operators are
calculated following the Rigorous Coupled-Wave Analysis (RCWA) method described
in~\cite{moharam1995}: the fields are expressed in terms of a Rayleigh expansion
in both homogeneous regions $z<0$ and $z>a$. In the corrugated region $0>z>a$, the
fields are developed in Fourier components. The Maxwell equations are solved in
each regions and writing the continuity of each Rayleigh
and Fourier components at the boundaries $z=0$ and $z=a$ leads to the reflection and
transmission coefficients for the grating. In the limit of an infinite number of
Fourier harmonics, this method solves exactly the diffraction of the fields by the
grating. Metallic gratings are known to be difficult to account for using the RCWA
method. We incorporate in the RCWA formalism the modifications presented in~\cite{lalanne1996}
which greatly improve the convergence rate for the reflection coefficients of a 
$p$-polarized light impinging on a metallic grating and our calculation
are performed with $N=51$ which shows converged results.

In the following, we apply formula~(\ref{eq:heat}) to compute the
heat transfer coefficient $h$ defined as $h=\frac{q}{T_{1}-T_{2}}$
for two temperatures $T_{1}$ and $T_{2}$ close enough to each other,
say for example $T_{1}=310$ K and $T_{2}=290$ K. We note that
$e_{T_1}-e_{T_2}$ acts as a cutoff
function for frequencies greater than the thermal frequency
$\omega_{T}=\frac{2\pi c}{\lambda_{T}}\approx 2.5\times
10^{14}$ rad\,s$^{-1}$ ($\lambda_{T} \approx 7.6$ $\mu$m). The
transmission factor $\mathcal{T}_{L}(\mathbf{k},\omega)$ thus
exhibits the mode structure for the problem under study
(Fig.~\ref{fig:gratingSchema}) while (\ref{eq:heat}) integrates the
contributions of all these modes to the heat transfer, taking into
account the values of their frequencies with respect to $\omega_{T}$
(more discussions below).

\begin{figure}[htbp]
  \begin{center}
    \includegraphics[width=8cm]{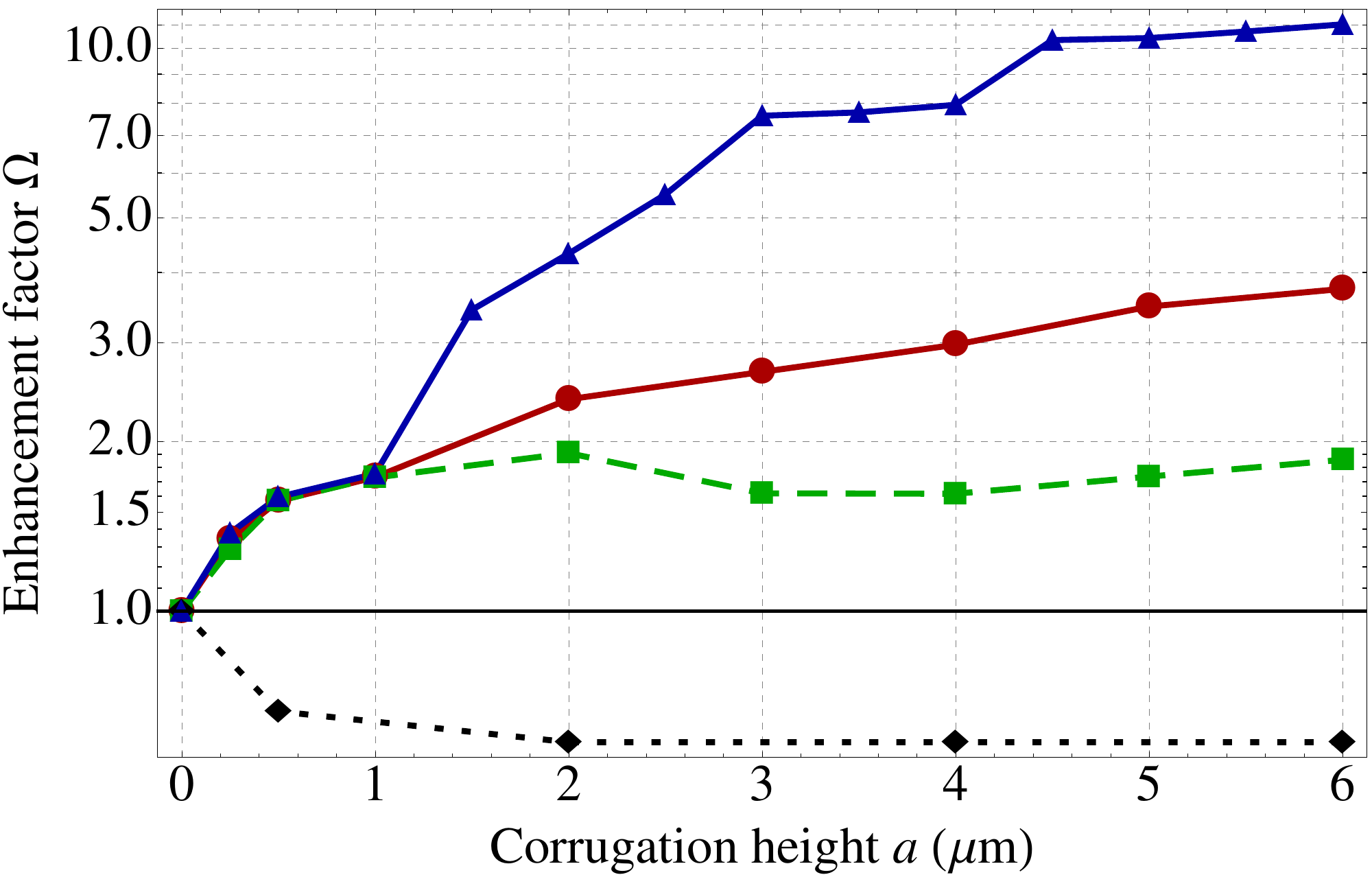}
  \end{center}
  \caption{The enhancement factor $\Omega$ between two gold gratings as a
  function of the depth $a$ of the corrugations, with the distance of
closest approach kept fixed $L=1$ $\mu$m. Red solid curve: period
$d=2.5$ $\mu$m. Green dashed curve: period $d=10$ $\mu$m. Black
dotted curve: proximity approximation. (colors online)}
  \label{fig:heat}
\end{figure}

For a depth of the corrugation $a=0$, we recover the heat transfer
coefficient $h_{0}(L)=0.16$ Wm$^{-2}$K$^{-1}$ between two gold plates
separated by a distance $L=1$ $\mu$m. For a non null depth $a$, we
introduce the factor of enhancement of heat transfer with respect to
non corrugated plates
\begin{equation}
  \label{eq:Exalt}
  \Omega=\frac{h(L)}{h_{0}(L)}.
\end{equation}
We present in Fig.~\ref{fig:heat} the enhancement factor $\Omega$ as
a function of the corrugation depth $a$, with the distance of
closest approach $L=1$ $\mu$m and the filling factor $p=0.5$ kept
fixed. The blue solid curve corresponds to a period $d=1$ $\mu$m for
the gratings while the red solid curve corresponds to a period $d=2.5$ $\mu$m.
The dashed curve corresponds to a period $d=10$
$\mu$m. As the corrugations become deeper, we see a striking
increase in the heat transfer coefficient. We note that the enhancement factor is
largely independent of the grating period up to a corrugation depth $a\approx 1$ $\mu$m. For a period $d=1$ $\mu$m for which the effect is more important, we get an enhancement up to a factor 10 for $a=6$ $\mu$m. For a period $d=2.5$ $\mu$m, the enhancement reaches nearly a factor 4 for $a=6$ $\mu$m. For the largest period $d=10$ $\mu$m, the enhancement still reaches nearly a factor 2 at $a=6$ $\mu$m. 

For comparison, we have shown as the dotted line in
Fig.~\ref{fig:heat} the prediction of the proximity approximation
(PA) which amounts to adding plane-plane heat transfer contributions,
as if they were independent,
\begin{equation}
  \label{eq:PA}
  \Omega^{PA}=p+(1-p)\,\frac{h_{0}(L+2a)}{h_{0}(L)}.
\end{equation}
As expected, the PA predicts a decrease of $\Omega$ when $a$ is
increased, in complete contradiction with the exact results shown by
the solid and dashed curves.

In the remainder of this letter, we analyze the electromagnetic mode
structure in order to explain the increase of the heat
transfer~\cite{biehs2010,pendry1983}. To this aim, we use the
scattering formula~(\ref{eq:heat}) and show that, as we increase the
corrugation depth, some modes of the system are indeed brought to
the infrared frequencies and thus are able to contribute to the heat
transfer. The mode structure is described by the transmission factor
$\mathcal{T}_L(\mathbf{k},\omega)$ which can reach its maximum value
1 at the resonances of the corrugated cavity. Our system is periodic
so that the mode structure, distributed over the whole range of
wavevectors in the absence of corrugations, now shows many branches
folded in the first Brillouin zone. More precisely, there are
$2(2N+1)$ branches where the factor 2 is due to the two
polarizations and the factor $2N+1$ is the number of orders (or
branches) used when taking into account mode coupling by diffraction
on the gratings.

\begin{figure}[htbp]
  \begin{center}
    \includegraphics[width=8cm]{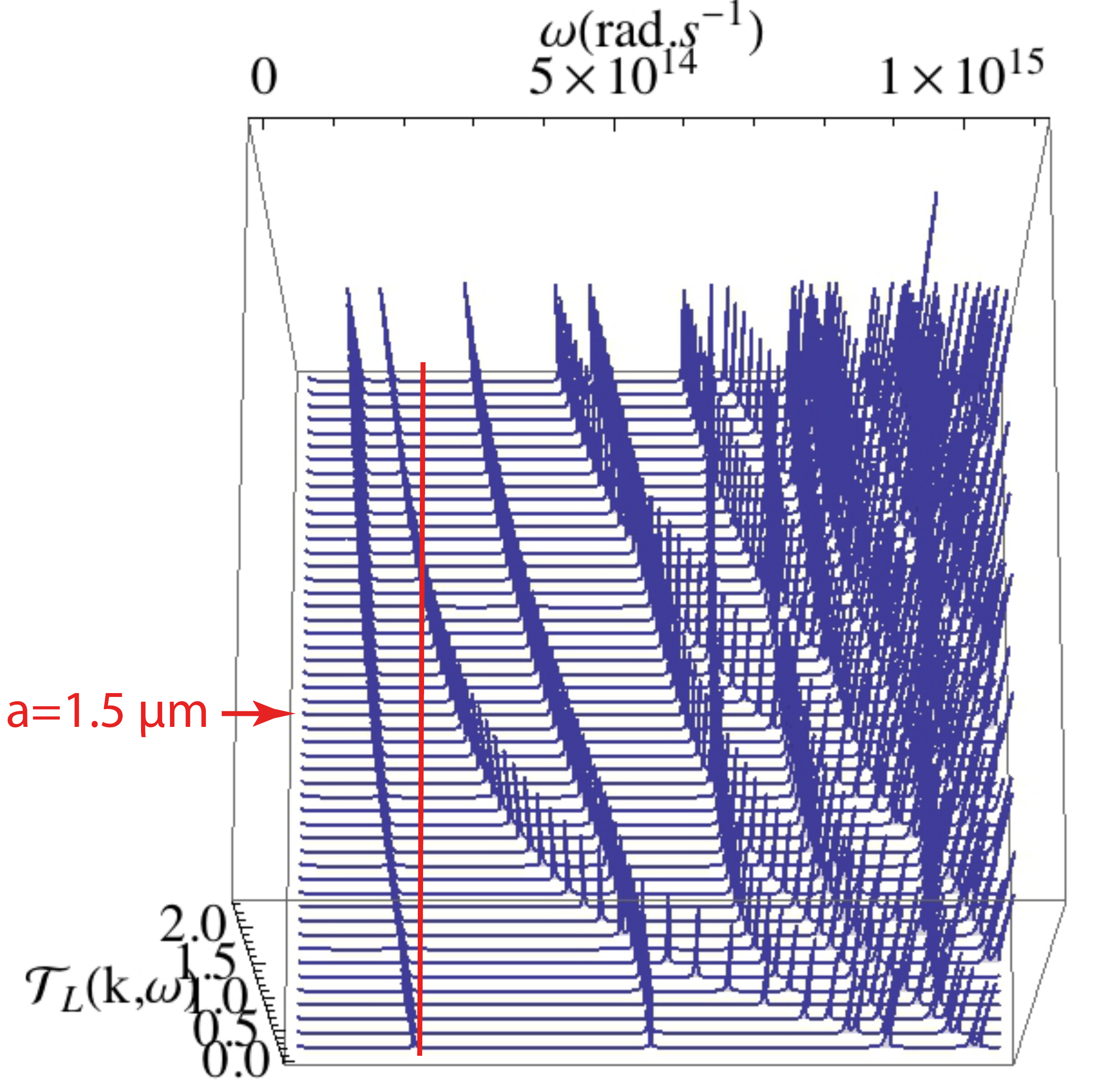}
  \end{center}
  \caption{The transmission factor for two gold gratings as a function of
  the frequency $\omega$ and the corrugations depth $a$. The lower curve
  is for plane-plane $a=0$ while the upper one is for a corrugations depth
  $a=3$ $\mu$m. The vertical red line is the light line. The horizontal arrow
  at $a=1.5$ $\mu$m shows a cut of this plot represented on
  Fig.~\ref{fig:compHugonin}. (colors online)}
  \label{fig:growingModes}
\end{figure}

We represent in Fig.~\ref{fig:growingModes} the sum of transmission
factors $\mathcal{T}_L(\mathbf{k},\omega)$ over all
polarizations and all branches. It is shown as a function of the
frequency $\omega$ and the depth of the corrugations $a$ for a fixed
value of the transverse wavevector
$\mathbf{k}=(\frac{\pi}{2d},0)$, here chosen to be in the
middle of the positive-$k_{x}$ first Brillouin zone. The plot
corresponds to the period $d=2.5$ $\mu$m, which was shown as the solid
curve in Fig.~\ref{fig:heat}. The vertical red line represents the
light line $\omega=ck_{x}\approx 1.88\times 10^{14}$ rad\,s$^{-1}$.

It clearly appears in Fig.~\ref{fig:growingModes} that the
transmission factor takes significant values only on resonances
which correspond to the mode structure of the corrugated cavity. The
transmission factor $\mathcal{T}_L(\mathbf{k},\omega)$ goes
to a maximum value of 1 for each non degenerate mode $(\mathbf{k},\omega)$; it 
can be 2 if two modes cross and we see one of these occurrences in the figure. 
The general trend is clear on
the diagram: as the depth $a$ of the corrugations is increased, new
modes appear, with frequencies decreasing as $a$ increases. When
these modes enter into the thermal window $\omega\lesssim\omega_{T}$
they contribute more and more to the heat transfer. This explains
the enhancement of the heat flux, due to the presence of additional
modes in the thermal window for a deeply corrugated structure.

We now examine in more detail the nature of the modes.
While varying the corrugation depth $a$ from 0 to 3 $\mu$m we can
follow the evolution of each mode. Note that, for $k_{y}=0$, the
polarizations $\sigma=s$ and $\sigma=p$ are not mixed (however, the
computation of $h$ takes into account all modes for which
polarization mixing is important).

\begin{figure}[hbtp]
  \begin{center}
    \includegraphics[width=8cm]{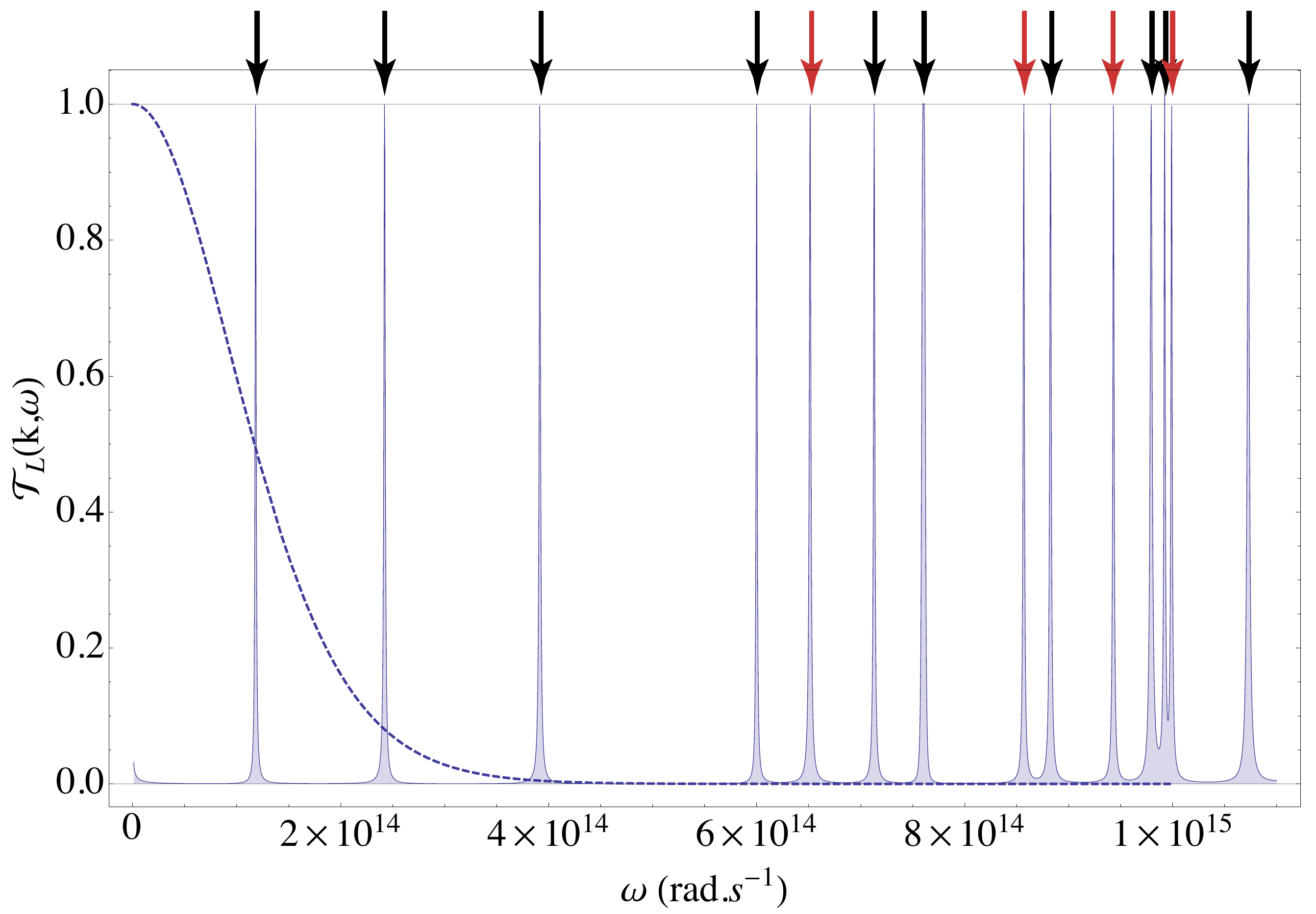}
  \end{center}
  \caption{The transmission factor for two gold gratings with corrugation
   depth $a=1.5$ $\mu$m as a function of frequency $\omega$. The arrows
   indicate the position of the modes in a direct mode calculation (red for
   $s$ polarization and black for $p$ polarization).
   The dashed curve is the function $\frac{e_{T_1}-e_{T_2}}{k_\mathrm{B}(T_1-T_2)}$.
   (colors online)}
  \label{fig:compHugonin}
\end{figure}

We show in Fig.~\ref{fig:compHugonin} the modes calculated for a
particular corrugation depth $a=1.5$ $\mu$m indicated by the red
horizontal line on Fig.~\ref{fig:growingModes}. The position of the
peaks have been confirmed through a direct mode
calculation~\cite{cao2002} of the eigenfrequencies of the structure
modes obtained for $p$ (black arrows) and $s$ (red arrows)
polarizations. In addition to the excellent agreement between the
peaks of the transmission factor and the directly calculated modes
(arrows on Fig.~\ref{fig:compHugonin}), direct mode calculations
show the fields and, therefore, allow us identifying the first few
modes. For the second $p$ polarization and the first $s$
polarization modes appearing at $\omega\approx 2.4\times
10^{14}$ rad\,s$^{-1}$ and $\omega\approx 6.5\times
10^{14}$ rad\,s$^{-1}$ in particular, the frequencies are largely
independent upon the value of $k_{x}$, which is usually the
signature of guided modes. By looking at the fields corresponding to
those two modes, we indeed confirmed that the electric field is to some
extent confined in the waveguides formed by the corrugations.

\begin{figure}[htbp]
  \begin{center}
    \includegraphics[width=8cm]{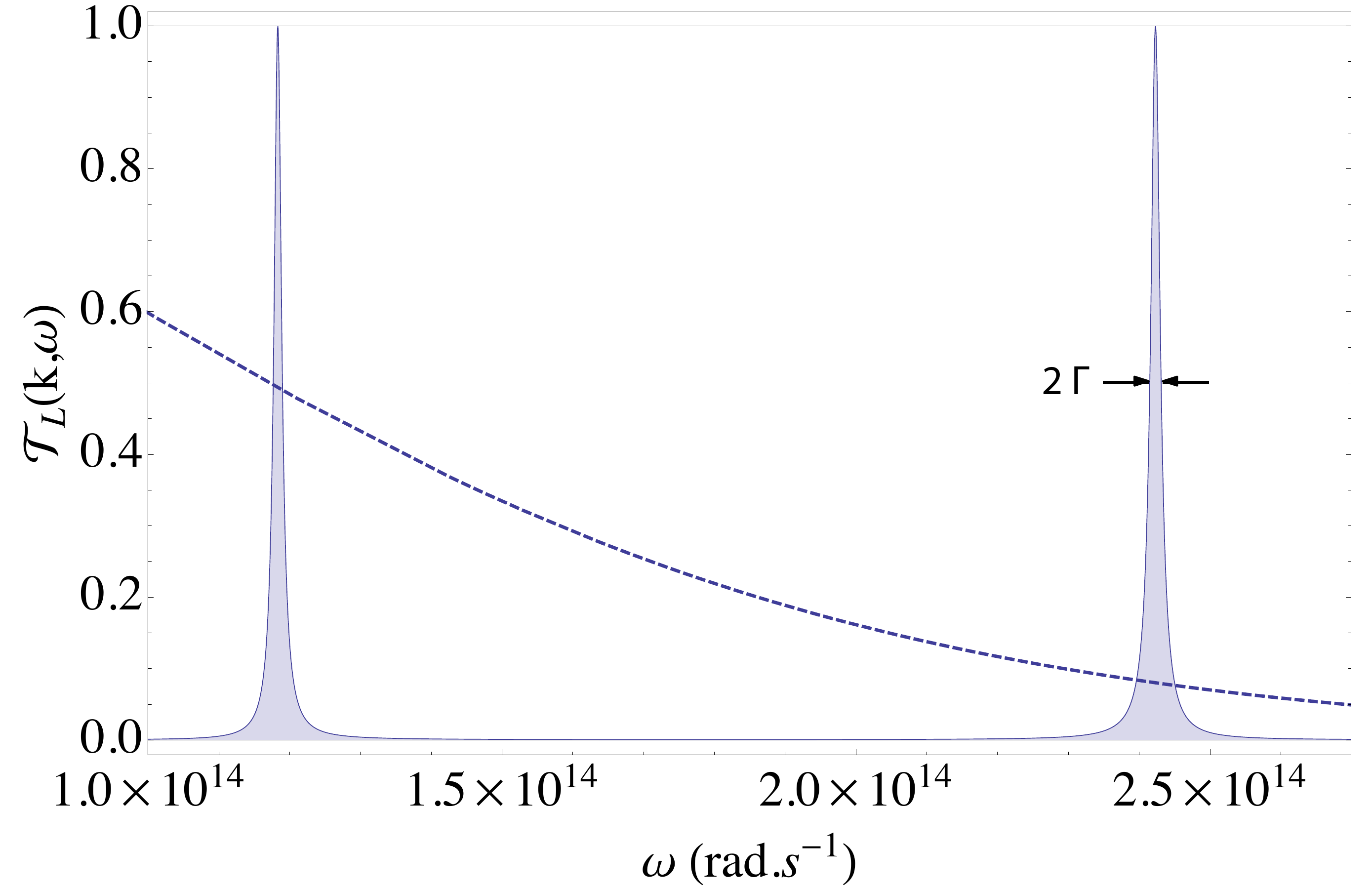}
  \end{center}
  \caption{Same as fig.~\ref{fig:compHugonin} for the first two modes which
  are in the thermal window.
  Each peak can be fitted by a Lorentzian of resonance frequency $\omega_{0}$ and
  half width at half maximum $\Gamma$. The dashed curve is the function
  $\frac{e_{T_1}-e_{T_2}}{k_\mathrm{B}(T_1-T_2)}$.}
  \label{fig:fineStructure}
\end{figure}

It is also worth discussing the shape of the resonance curve drawn
by the variation of the transmission factor in the vicinity of a
mode. In fig.~\ref{fig:fineStructure}, we focus on the modes which
lie inside the thermal window. In the case considered here of sharp,
isolated modes, the resonance of the transmission factor shows a
Lorentzian profile. We have checked that the two parameters of this
profile are identified respectively to the real and imaginary parts
of the complex frequency, with mode calculation of the dissipative
structure defined with complex frequencies and real
wavevectors~\cite{joulain2005}. This proves that the variation of
the transmission factor contains all the relevant information about
the mode structure. Not only the frequencies but also their finite
lifetime are well described in the case considered here of lossy
materials.

This discussion allows one predicting the effect of a change of the
dissipation parameter $\gamma$. As this parameter is the only one to
determine the widths of the peaks in the transmission factor
$\mathcal{T}_L(\mathbf{k},\omega)$, one deduces that these widths
vary linearly with $\gamma$. As a direct consequence of
(\ref{eq:heat}) and as long as the modes remain sharp and
isolated, it follows that the heat fluxes vary in proportion of
$\gamma$, so that the enhancement factor $\Omega$, defined in
(\ref{eq:Exalt}) and drawn on Fig.~\ref{fig:heat}, is independent
of the dissipation parameter $\gamma$.

We have theoretically demonstrated the enhancement of the heat
transfer between two nanocorrugated gold plates in comparison with
flat plates with the same distance of closest approach. This
enhancement is due to the presence of additional modes in the thermal
frequency window contributing to the heat transfer. We have
described all the relevant information about the mode structure in
terms of the transmission factor $\mathcal{T}_L(\mathbf{k},\omega)$
which appears in the scattering formula for the heat flux. We have
discussed the enhancement of the heat transfer in a regime where the
three characteristic lengths of the problem (the distance $L$
between the gratings, the period $d$ of the gratings and the height
$a$ of the corrugations) are of the same order. We stress that
neither the proximity nor the effective medium approximations can
work in this regime. We have in fact shown that the proximity
approximation predicts a decrease of the heat transfer, in
complete contradiction with the striking enhancement of the heat flux
observed in the exact results.

\begin{acknowledgments}
The authors thank the ESF Research Networking Programme CASIMIR
(www.casimir-network.com) for providing excellent possibilities for
discussions and exchange. The research described here has been
supported by Triangle de la Physique contract EIEM 2010-037T. This
work was carried out under the auspices of the National Nuclear
Security Administration of the U.S. Department of Energy at Los
Alamos National Laboratory under Contract No. DE-AC52-06NA25396. RG
and DARD thank LANL and ENS, respectively, for funding their stay at
these institutions, where part of this work was done.
\end{acknowledgments}


\providecommand{\noopsort}[1]{}\providecommand{\singleletter}[1]{#1}%

\end{document}